\documentclass[superscriptaddress, showpacs, twocolumn,nofootinbib,preprintnumbers, letterpaper,accepted=2019-01-14]{quantumarticle}
\pdfoutput=1

\usepackage{amsmath,amssymb,amsfonts}
\usepackage{mathtools}
\mathtoolsset{centercolon}
\usepackage{tikz}
\usetikzlibrary{calc,positioning,arrows,shapes.symbols,automata,chains,fit,shapes,er,petri,patterns}

\usepackage[utf8]{inputenc}
\usepackage[english]{babel}
\usepackage[T1]{fontenc} 

\usepackage[colorlinks=true]{hyperref}

\usepackage[numbers,sort&compress]{natbib}

\DeclarePairedDelimiter{\abs}{\lvert}{\rvert}

\renewcommand{\(}{\left(}
\renewcommand{\)}{\right)}
\newcommand{\lket}{\left|}
\newcommand{\rket} {\right\rangle}
\newcommand{\microspace}{\mspace{0.5mu}}
\newcommand{\ket}[1]{\lket\microspace #1 \microspace\rket}

\newcommand{\I}{\mathcal{I}}

\newcommand{\ip}[2]{\left\langle #1 , #2\right\rangle} 

\begin{document}

\title{POVMs are equivalent to projections for perfect state exclusion of three pure states in three dimensions}
\date{\today}
\author{Abel Molina}
\affiliation{Institute for Quantum Computing and School for Computer Science, University of Waterloo}
\orcid{0000-0003-2618-3071}

\begin{abstract}

Performing perfect/conclusive quantum state exclusion means to be able to discard with certainty at least one out of $n$ possible quantum state preparations by performing a measurement of the resulting state. This task of state exclusion has recently been studied at length in \cite{bandyopadhyay2014conclusive}, and it is at the heart of the celebrated PBR thought experiment \cite{pusey2012reality}. When all the preparations correspond to pure states and there are no more of them than their common dimension, it is an open problem whether POVMs give any additional power for this task with respect to projective measurements. This is the case even for the simple case of three states in three dimensions, which is mentioned in \cite{caves2002conditions} as unsuccessfully tackled. In this paper, we give an analytical proof that in this case considering POVMs does indeed  not give any additional power with respect to projective measurements. To do so, we first make without loss of generality some assumptions about the structure of an optimal POVM. The justification of these assumptions involves arguments based on convexity, rank and symmetry properties. We show then that any pure states perfectly excluded by such a POVM meet the conditions identified in \cite{caves2002conditions} for perfect exclusion by a projective measurement of three pure states in three dimensions. We also discuss possible generalizations of our work, including an application of Quadratically Constrained Quadratic Programming that might be of special interest.
\end{abstract}

  \maketitle

\section{Context and Motivation}

The task of quantum state exclusion corresponds to a setting where an agent Alice is given a quantum system. The state of this system is chosen at random between $n$ options $\{\rho_1, \ldots, \rho_n\}$, with corresponding non-zero probabilities $\{p_1, \ldots , p_n \}$. It is unknown to Alice which of the $\rho_i$ was chosen, but she does know the $\{\rho_i\}$ and $\{p_i\}$ values characterizing the corresponding distribution. Alice's goal in the state exclusion task is to be able to give an index $j$ such that the state was \emph{not} prepared in the state  $\rho_j$. When Alice can achieve this with probability $1$, we will say that we have \emph{perfect} state exclusion. This task of state exclusion has recently been studied at length in \cite{bandyopadhyay2014conclusive},  and is at the heart of the celebrated PBR thought experiment \cite{pusey2012reality},  where  \cite{caves2002conditions} (the article from where we take the problem we solve) is credited as the original source for the concept. The concept of this task has also been used for proving results in the context of quantum communication complexity \cite{perry2015communication, liu2016doubly}, as well as for designing quantum signature schemes \cite{arrazola2015multiparty}.

Formalizing further this concept of state exclusion,  \cite{bandyopadhyay2014conclusive} obtains the following semidefinite programming (SDP) formulation:

\vspace{-4mm}
\begin{center}
    \begin{equation}
   \label{eq:GeneralSDP}
    \begin{aligned}
      \text{minimize:}\quad & \sum_i  p_i \ip{M_i}{\rho_i} \\
      \text{subject to:}\quad & \sum_i{M_i} = \I  \\
          & M_i  \geq 0.
    \end{aligned}
    \end{equation}
\end{center}

\noindent where $M_i  \geq 0$ means that $M_i$ is positive semi-definite. Being able to perform perfect state  exclusion corresponds to the optimal value of this SDP being equal to $0$. Similarly, any optimal solution to the semidefinite program corresponds to an optimal positive-operator valued measure (POVM)  for state exclusion. Note that since we are only concerned with perfect state exclusion, we can just ignore the $p_i$ in the rest of this presentation, since whether the value of the SDP is $0$ or not does not depend on them.

Perfect exclusion of quantum states is also a meaningful concept in the context of the foundations of quantum mechanics, in particular when considering the topic of quantum state compatibility.  In that framework, one considers several quantum states $\{ \rho_1 , \ldots, \rho_n \}$ as different beliefs about the same system. Then, one can ask whether the outcome of a measurement on the system will disprove some of these beliefs, or they will all still be possible. In the latter case, we say that the states are \emph{compatible} with each other. Different ways of formalizing this idea will lead to different definitions of quantum state compatibility.  \cite{caves2002conditions}  proposes several formalizations, one of which corresponds to the impossibility of performing perfect state exclusion. Since this formalization is a generalization of previous work by Peierls \cite{peierls1991more}, they refer to it as post-Peierls (PP) compatibility.

In more detail, the post-Peierls compatibility of several quantum states $\{ \rho_1 , \ldots, \rho_n \}$ (relative to a subset $S$ of all POVMs) means that for all measurements in $S$, there will be at least one outcome that can be obtained with non-zero probability for all of the possible states/beliefs $\{ \rho_1 , \ldots, \rho_n \}$. If we consider the negation of this definition, we obtain that this negation corresponds with the existence of a measurement in $S$ such that each outcome of the measurement excludes at least one of the quantum states, which corresponds to an agent being able to perform perfect state exclusion given a mixture of the quantum states  $\{ \rho_1 , \ldots, \rho_n \}$ and access to measurements in $S$. When the set $S$ of allowed measurements corresponds to the set of all POVMs, the corresponding compatibility criteria is called PP-POVM compatibility. When $S$ is restricted to the set of projective measurements (or more precisely, the set of measurements defined by one-dimensional orthogonal projectors), \cite{caves2002conditions} names the corresponding criteria as PP-ODOP compatibility.

One can consider the case where all of the $n$ states/beliefs  $\{ \rho_1 , \ldots, \rho_n \}$ are known to belong to a particular subset $A$ of all quantum states, and ask whether for all such tuples of $n$ beliefs in $A$ the PP-ODOP and PP-POVM criteria will coincide with each other. When this happens, we will say that in that context PP-ODOP=PP-POVM. Note that this is equivalent to projections being optimal for perfect state exclusion within the context of input states in $A$.

In \cite{caves2002conditions}, the authors identify a necessary and sufficient condition for PP-ODOP incompatibility of $3$ pure states $\{a, b, c\}$ in $3$ dimensions (i.e. they establish a condition for the states to be perfectly  excludable via a projective measurement). This condition can be expressed in terms of the magnitudes of their inner products, given by  $\abs{\ip{a}{b}}, \abs{\ip{a}{c}}, \abs{\ip{b}{c}}$, and which we will denote as $j_1$, $j_2$ and $j_3$, respectively.  In particular, the condition obtained in   \cite{caves2002conditions}   is that  $3$ pure states will be PP-ODOP incompatible whenever

\vspace{-4mm}

\begin{align}
j_1^2 + j_2^2 + j_3^2 + 2j_1 j_2 j_3  \leq 1 \label{Caves-Fuchs-SchackInequality}.
\end{align}

\noindent We will refer to this formula as the \textbf{Caves-Fuchs-Schack inequality}, after the authors of \cite{caves2002conditions}. Note that we have corrected in our presentation of this formula the original strict inequality sign that they use, following the indications in \cite{barrett2014no, stacey2016sic}, and we have also merged the two conditions from the original presentation in \cite{caves2002conditions} into one single condition.

When this result was introduced in \cite{caves2002conditions}, it was mentioned that the authors were not able to prove that  $\text{PP-ODOP} = \text{PP-POVM}$ in the context of $3$ pure states in $3$ dimensions, despite having numerical evidence that this is the case. The authors also present results establishing that this is the first open case  -- they cite previous work \cite{mermin2001whose} showing that for two pure states in any dimension $\text{PP-ODOP} = \text{PP-POVM}$, and establish that for $k > 2$ pure states in $2$ dimensions this will not necessarily be the case.

We will give now an analytical proof which answers the corresponding question, by showing that $\text{PP-ODOP} = \text{PP-POVM}$ in the context of $3$ pure states in $3$ dimensions.

Our work can be seen as part of the line of work that studies POVMs in the context of low-dimensional systems of a fixed dimension. For example, \cite{vertesi2010two} and \cite{PhysRevA.96.022312} recently examined 2-dimensional POVMs in the contexts of nonlocal games and quantum state discrimination, respectively, while \cite{zhu2010two} looked into 4-dimensional POVMs in the context of imposing symmetry conditions.

We conclude with a discussion about different ways in which our work might be generalized. Of special interest here might be our discussion on the usage of Quadratically Constrained Quadratic Programming (QCQP) to model the $n$-dimensional variant of the question we solve. This is a type of mathematical optimization formalism that has seen a  large number of applications in recent years, but only limited usage so far within the context of quantum information processing. To our knowledge, this is the first time that state exclusion of pure states through projections is expressed through a problem in a standard form of a mathematical optimization framework. 

In our derivation, we will use standard quantum information theory notation and vocabulary -- standard texts on the topic (e.g. \cite{watrous2011theory, wilde2013quantum}) can be consulted for definitions of the corresponding terms.

\section{Main derivation}

\subsection{Restrictions that can be imposed  without loss of generality on POVMs that achieve perfect exclusion}

Our goal is to prove that for any set of 3 pure states in $3$ dimensions that are perfectly excluded by a POVM, they are also perfectly excluded by a projective measurement. Following Equation~\eqref{eq:GeneralSDP} and our analysis of it, we can identify the perfect exclusion of three pure states $a, b \text{ and } c$ with obtaining an optimal value of $0$ in the following semidefinite program:

\vspace{-4mm}
\begin{center}
    \begin{equation}
   \label{eq:GeneralSDP3}
    \begin{aligned}
      \text{minimize:} \quad & a^*M_1 a +  b^*M_2b + c^*M_3c \\
      \text{subject to:}\quad & \sum_i{M_i} = \I  \\
          & M_i  \geq 0,
    \end{aligned}
    \end{equation}
\end{center}

\noindent Note that all the operators involved can be represented as $3 \times 3$ matrices. It is well-known in convex optimization that the solution to optimizing a linear function over a non-empty compact convex set in a finite dimensional Hilbert space can be assumed without loss of generality to be an extreme point of the set of feasible solutions\footnote{This fact follows from applications of the Krein-Milman and Extreme Value theorems, which in their most general versions require in fact constraints less strong than the ones we have here.}  (note that in this case, that feasible set is the set of POVMs). Therefore, we can assume that at most one out of the $M_i$ has rank greater than $1$. Otherwise, assume for the sake of contradiction that two of them (say $M_1$ and $M_2$) have rank at least $2$, so there is a common vector $u$ in the images of $M_1$ and $M_2$. Then, for $\epsilon$ small enough both $\{M_1 + \epsilon uu^*, M_2 - \epsilon uu^*, M_3 \}$ and $\{M_1 - \epsilon uu^*, M_2 + \epsilon uu^*, M_3 \}$  are POVMs, which implies $\{M_1, M_2, M_3\}$ is not an extreme point of the feasible set.

Without loss of generality, we can permute indices so that the ranks of  $M_1$, $M_2$, and $M_3$ are sorted in non-increasing order. Also, if $M_1$ has rank $3$ it cannot exclude any quantum state, so it must be the case that its rank is at most $2$.  Note too that if the ranks are of the form $(1,1,1)$, it is not hard to see that the condition $M_1 + M_2 + M_3 = \I$ implies that $\{M_1, M_2, M_3\}$ form a projective measurement themselves. Similarly, in the case where the ranks are of the form $(2,1,0)$, $M_1 = \I - M_2$ implies that $M_1$ and $M_2$ form a projective measurement (this is because for the right hand side $\I - M_2$ to have rank $2$, $M_2$ must have its non-zero eigenvalue equal to $1$, which implies then the same for the eigenvalues of $\I - M_2 = M_1$).

We can assume then without loss of generality that there is an optimal POVM $\{ M_1, M_2, M_3 \}$ with ranks of the form $(2,1,1)$. We can also choose now without loss of generality  to work in a basis such that $M_1$ is diagonal and it perfectly excludes~$a=\ket{0}$.

We have then that $M_1$ will be determined by the choice of a real diagonal vector $(0, 1-x, 1-y)$, and $M_2$ and $M_3$ by a choice of complex vectors $v= (v_1, v_2, v_3)$ and $w = (w_1, w_2, w_3)$ such that $M_2 = vv^*$ and $M_3 = ww^*$. We claim now that we can assume $y=0$, $v_1 \neq 0$, $w_1 \neq 0$, $v_2\neq 0$, $w_2 \neq 0$, $v_3=0$, $w_3=0$ To see why, consider the following five observations:

\vspace{1cm}
\begin{enumerate}

\item The condition $M_1 + M_2 + M_3 = \I$ corresponds to the equations

\begin{align}
v_1v_1^* + w_1w_1^* & = 1 \label{firstInSys}\\
v_2v_2^* + w_2w_2^* & = x  \label{secondInSys}\\
v_3v_3^* + w_3w_3^* & =  y  \label{thirdInSys}\\
v_1v_2^*  & =  -w_1w_2^*  \label{fourthInSys} \\
v_1v_3^*  & =  -w_1w_3^*  \label{fifthInSys} \\
v_2v_3^*  & =  -w_2w_3^*.  \label{sixthInSys}
\end{align}

\item  We can assume $v_1 \neq 0$ and $w_1 \neq 0$, as otherwise $\{M_1, M_2, M_3\}$ can be trivially transformed into a projective measurement. To see this, suppose for example that $w_1 = 0$. Then,~\eqref{firstInSys} implies that $\abs{v_1} = 1$,~\eqref{fourthInSys} that $v_2=0$, and~\eqref{fifthInSys} that $v_3=0$.  We have then that $M_2$ is diagonal, and its diagonal is equal to $(1, 0, 0)$. This implies that $M_3$ is diagonal as well, while $w_1=0$ implies that the first term in its diagonal is equal to $0$, so we can group $M_1$ and $M_3$ into a single operator and obtain a projective measurement.  

\item Suppose we had $v_2 \neq 0$ and $v_3 \neq 0$. Then,~\eqref{sixthInSys} implies $w_2 \neq 0$ and $w_3 \neq 0$. This means we can divide~\eqref{fourthInSys} by~\eqref{fifthInSys} and the conjugate of~\eqref{sixthInSys}, and obtain that $\frac{v_1}{w_1} = \frac{v_2}{w_2} = \frac{v_3}{w_3}$. Let $\lambda$ be the value of these ratios. Then, each of equations~\eqref{fourthInSys}-\eqref{sixthInSys} implies that $\lambda=0$, which contradicts $v_1 \neq 0$.

\item We have then that either $v_2=0$ or $v_3=0$, and by symmetry we can assume without loss of generality that $v_3=0$. Then, $w_3= 0$ as well, since otherwise~\eqref{fifthInSys} would imply $w_1=0$, which we know not to be the case.  \eqref{thirdInSys} implies then that $y=0$.  

\item If we were now to additionally impose that $v_2 = 0$, \eqref{fourthInSys} would imply that $w_2=0$, which can only happen when $x=0$, by \eqref{secondInSys}. However, in the $x=0$ case we have that $M_1$ is a projection on $\ket{1}$ and $\ket{2}$, and $M_2$ and $M_3$ can be merged into a projection on $\ket{0}$, so there trivially is an optimal projection for state exclusion, and the case is not of interest to us. We can assume then that $v_2 \neq 0$, and similarly that $w_2 \neq 0$.

\end{enumerate}

We can introduce now a parameter $r$, which determines the distribution of the weight $x$ between $M_2$ and $M_3$, and let $\abs{v_2}^2$ be equal to $ x \frac{1}{r + 1}$ ($r \in (0, \infty)$). We have then that \eqref{secondInSys} implies $\abs{w_2}^2 = x \frac{r}{r + 1}$, and that \eqref{fourthInSys} and \eqref{firstInSys} imply then that $\abs{v_1}^2 = \frac{r}{r+1}$, $\abs{w_1}^2 = \frac{1}{r+1}$. The magnitudes of each element of $v$ and $w$ are then completely characterized by the values of $r$ and $x$.

As for the phases of the elements of $v$ and $w$, we can assume that $v_1, w_1 \in \mathbb{R}$ without affecting the values of $M_2$ and $M_3$. Then, if the phase of $v_2$ is given by $\theta$,  \eqref{fourthInSys} implies that the phase of $w_2$ is given by $\pi + \theta$. 

We reach then our final form for what a POVM $\{M_1, M_2, M_3\}$ for perfect state exclusion of $3$ pure states in $3$ dimensions can be assumed to be without loss of generality. In matrix form, it is given by

\begin{align}
M_1 & = \begin{pmatrix}
0 & 0 & 0 \\
0 & 1 - x & 0 \\
0 & 0 & 1 \\
\end{pmatrix}, \label{firstMatrixEquation} \\
M_2 & =  \frac{1}{r+1}\begin{pmatrix}
r & e^{-i \theta} \sqrt{rx}& 0 \\
e^{i \theta}\sqrt{rx} & x & 0 \\
0 & 0 & 0  \label{secondMatrixEquation} \\
\end{pmatrix} \\
M_3 & = \frac{1}{r+1}  \begin{pmatrix}
1 & -e^{-i \theta}\sqrt{rx} & 0 \\
-e^{i \theta}   rx  &  rx & 0 \\
0 & 0 & 0 \\
\end{pmatrix} \label{thirdMatrixEquation} 
\end{align}

\noindent where $0 < x < 1$, $r \in (0, \infty)$, $0 \leq \theta < 2\pi$.

\subsection{Verification that any states perfectly excluded by our parametrized optimal POVM satisfy the Caves-Fuchs-Schack~inequality}

We look first at the structure of the states $b$ and $c$ perfectly excluded by $M_2$ and $M_3$, and obtain that it is enough to consider a one-parameter family for each of them. Let $b$ be given by $(b_1,b_2,b_3)$, and $c$ by $(c_1,c_2,c_3)$. Then, our conclusion follows from the following five observations:

\begin{enumerate}
 
\item As usual, we can get rid of unphysical global phases, and assume $b_1$ is a real positive number. This is because multiplying $b$ by a phase   will not affect the value of our semidefinite program \eqref{eq:GeneralSDP3}, and it will not affect either the satisfaction of the Caves-Fuchs-Schack inequality.

\item  It can be seen from \eqref{secondMatrixEquation} that the value of $b_2$ is completely determined by the value of $b_1$ by the constraint $M_2 b = 0$ (which is equivalent to $b^* M_2 b = 0$, since $M_2$ is Hermitian). In particular, one obtains that $b_2 = - b_1 e^{i \theta}\sqrt{\frac{r}{x}}$.
 
\item The fact that $b$ has norm $1$ (since it represents a pure state) allows us now to express the magnitude of $b_3$ as a function of $b_1$. In particular, the magnitude of $b_3$ is given by $\sqrt{1 - b_1^2 \(1 +  \frac{r}{x} \) }$, while its phase, which we will denote by $\vartheta$, can take any value.

Note that this implies an upper bound on $b_1^2$, given by $1/ \(1 + \frac{r}{x} \) $.
  
\item A similar analysis applies to $c$, and we have that it can be parametrized by a real positive value $c_1$ such that $0 \leq c_1^2 \leq  1/ \(1 + \frac{1}{rx} \) $, together with the phase $\gamma$ of $c_3$. In this case, the value of $c_2$ is given by $c_1 e^{i \theta}\sqrt{ \frac{1}{rx}}$, and the magnitude of $c_3$ is given by $\sqrt{1 - c_1^2 \(1 +  \frac{1}{rx} \) }$.

\item We can assume now that the phases  $\vartheta$ and $\gamma$  of $b_3$ and $c_3$ are selected in order to maximize the left hand side of the Caves-Fuchs-Schack inequality. The reason we can do this is because we are interested in proving that the Caves-Fuchs-Schack inequality holds, so this is a worst-case scenario in our situation.

To do so, note that $j_1=b_1$ and $j_2 = c_1$, so they do not depend on the phases of $b_3$ and $c_3$. Therefore, maximizing the left hand side of the Caves-Fuchs-Schack inequality will be equivalent to maximizing $j_3 = \abs{b^*c}$. To do that, we compute first the value of $b^*c$, given by

\begin{align}
 & e^{i(\gamma - \vartheta)} \sqrt{1 - b_1^2 \(1 + \frac{r}{x}\)}\sqrt{1 - c_1^2 \(1 + \frac{1}{rx}\)}   \nonumber \\
 + ~ & b_1 c_1 - \frac{1}{x} b_1 c_1.
\end{align}

The magnitude of this expression will be the largest possible  whenever the term in the first line interferes constructively with the term in the second line. This will happen whenever the first line term is also real, and has the same sign as $b_1 c_1 (1 - 1/x)$  We can in fact achieve this by picking $\gamma = \vartheta + \pi$, since $0 <x < 1$. We obtain then that in our worst-case situation, 

\begin{align}
j_3=&  \sqrt{1 - b_1^2 \(1 + \frac{r}{x}\)}\sqrt{1 - c_1^2 \(1 + \frac{1}{rx}\)}   \nonumber  \\
 & + b_1c_1 (1/x -1)  \label{j3given}.
\end{align}

 \end{enumerate}

\noindent 

\noindent The Caves-Fuchs-Schack inequality is expressed then in our case as 

\vspace{-5mm}
\begin{align}
\label{eq:finalFormulation}
j_3^2 + b_1^2 + c_1^2 + 2j_3b_1c_1 \leq 1,
\end{align}

\noindent where  $j_3$ is given in \eqref{j3given}, $ x~\in~(0, 1)$,   $r~\in~(0, \infty)$, $b_1~\in~[0, \sqrt{1/ \(1 + \frac{r}{x} \) })$, $c_1~\in~[0, \sqrt{ 1/ \(1 + \frac{1}{rx} \)})$. We will refer from now on to the left hand side of~\eqref{eq:finalFormulation}  as $f(x, r,b_1, c_1)$. If $b_1=0$ or $c_1=0$, a simple algebraic manipulation of the value of $j_3$ gives us that $f(x, r,b_1, c_1)  \leq 1$. Expanding the value of $j_3$, we have that $f(x, r,b_1, c_1)$ is given by

\begin{align*}
& ~b_1^2 c_1^2 (1 + 1/x^2 -2/x) \\
& + \(1-b_1^2\(1+\frac{r}{x}\)\)\(1-c_1^2\(1+\frac{1}{rx}\)\)   \\
& + 2b_1c_1 (1/x - 1)\sqrt{1-b_1^2\(1+\frac{r}{x}\)}\sqrt{1-c_1^2\(1+\frac{1}{rx}\)} \\
& + b_1^2 + c_1^2 + 2b_1^2c_1^2(1/x - 1) \\
& + 2b_1c_1\sqrt{1-b_1^2\(1+\frac{r}{x}\)}\sqrt{1-c_1^2\(1+\frac{1}{rx}\)} \\
& = 1  -b_1^2 \frac{r}{x}  - c_1^2 \frac{1}{rx} + c_1^2b_1^2 \( \frac{2}{x^2 }+ \frac{1}{x} \(r + \frac{1}{r} \) \) \\ 
& + 2b_1c_1 \frac{1}{x}\sqrt{1-b_1^2\(1+\frac{r}{x}\)}\sqrt{1-c_1^2\(1+\frac{1}{rx}\)}.
\end{align*}

To prove that this is less or equal than $1$, one can act similarly to the standard proof for $x + \frac{1}{x} \geq 2$, moving everything to one side of the inequality and writing as a square what one obtains. In more detail, multiplying by $x$ and dividing by $b_1^2c_1^2$ our last expression, we have that $f(x, r,b_1, c_1)$ will be less or equal than $1$ whenever 

\begin{align}
2\sqrt{\frac{1}{b_1^2}-\(1+\frac{r}{x}\)}\sqrt{\frac{1}{c_1^2}-\(1+\frac{1}{rx}\)} \nonumber \\
\leq r \frac{1}{c_1^2}  + \frac{1}{r} \frac{1}{b_1^2}  - \(\frac{2}{x} + \(r + \frac{1}{r} \)  \)  \label{endingOfXAppearance}
\end{align}

Observe now that both sides of this inequality are positive. This is trivial for the left hand side, and follows for the right hand side from the previous obtained upper bounds on $b_1$ and $c_1$. If we square both sides of this inequality and simplify the resulting expression, we obtain

\vspace{-5mm}

\begin{align}
\Big( r^2 \( \frac{1}{c_1^4} - \frac{2}{c_1^2} + 1 \) + \frac{1}{r^2}\(\frac{1}{b_1^4} - \frac{2}{b_1^2} + 1 \) \nonumber \\
+ 2\( \frac{1}{c_1^2} + \frac{1}{b_1^2} -\frac{1}{b_1^2c_1^2} -1 \)  \Big) \geq 0 \label{xNotThereAnymore}.
\end{align}

\noindent This can be rewritten as

\begin{align}
\( r\(\frac{1}{c_1^2} - 1 \) - \frac{1}{r}\(\frac{1}{b_1^2} -1\) \)^2 \geq 0,
\end{align}

\noindent which is true, so we have successfully proved that $a$, $b$ and $c$ satisfy the Caves-Fuchs-Schack inequality, and therefore can be excluded by a projective measurement.

Note that $x$ is not involved at all in \eqref{xNotThereAnymore}, although one can verify computationally that the difference between the left hand side and the right hand side of \eqref{endingOfXAppearance} does depend on $x$.

\section{Perspectives for generalization}

\subsection{Usage of Quadratically Constrained Quadratic Programs (QCQPs)}

\label{subsec:QCQP}

We will now discuss how to study the perfect exclusion of $n$ pure states by a projection through a  collection of Quadratically Constrained Quadratic Programs (QCQPs). For a situation with a $n$-dimensional complex variable $x$ and $m$ constraints, the standard form for such a program can be taken to be

\vspace{-6mm}
\begin{center}
    \begin{equation}
   \label{eq:generalQCQP}
    \begin{aligned}
      \text{minimize: }\quad &  x^*Gx \\
      \text{subject to:} \quad & x^* C_k x \geq l_k, \forall k \in \{1 , \ldots, m\},\\
    \end{aligned}
    \end{equation}
\end{center}

\noindent where the $l_k$ take real values, and $G$ and the $C_k$ are $n \times n$ Hermitian matrices.

This is a type of mathematical optimization formalism that has received considerable attention in recent years, with  wide-ranging applications in science and engineering (see \cite{aholt2012qcqp, huang2014randomized, li2014special, bose2015quadratically} for just a few amongst many relevant examples). There has also been a considerable number of results about the theoretical structure of the corresponding problems and the design of algorithms that can solve them (see e.g. \cite{konar2015hidden, josz2015moment, park2017general}).  However, there have only been a handful of applications so far \cite{liang2007bounds, fan2008improved, audenaert2009quantum, thiang2010some} of the QCQP model to quantum information processing.

In our collection of QCQPs, there will be one program for every $n$-combination with repetition $\{s_1, \ldots, s_n\}$ out of the set $\{w_1, \ldots, w_n\}$ of states to be excluded. Each choice represents a possibility for how the states excluded after obtaining different outcomes of the projection relate to each other, and the reason why we need to consider those choices is that two different outcomes of the projection could plausibly lead to excluding the same state (which in the POVM case would be handled by grouping those two outcomes into the same one). In particular, each of the corresponding QCQPs for perfect state exclusion via projections formalizes the following two ideas:
\begin{itemize}
\item A projection in $n$ dimensions corresponds to a choice of $n$ unit vectors $\{v_1, \ldots, v_n\}$ that are pairwise orthogonal.
\item We would like for every $v_i$ to be orthogonal to the corresponding $s_i$.
\end{itemize}

\vspace{5cm}

These ideas are then reflected in the following QCQP:

\vspace{-6mm}
\begin{center}
    \begin{equation}
   \label{eq:QCQP}
    \begin{aligned}
      \text{minimize: }\quad & \sum_i   v_i^*(s_is_i^*)v_i \\
      \text{subject to:} \quad & \\
          & v_i^*v_j + v_j^*v_i  = 0 \\
          & \quad \quad  \forall {i,j \in \{1, \ldots, n\}} \ s.t. \ i < j,\\
          & v_i^*v_i  = 1, \forall i \in \{1 , \ldots, n\},\\
          & v_i \in \mathbb{C}^n , \forall i \in \{1 , \ldots, n\}.
    \end{aligned}
    \end{equation}
\end{center}

Note that we have written $v_i^*v_j + v_j^*v_i  = 0$ rather than $v_i^*v_j =0$, in order to have the matrix representing the constraint be Hermitian, as required in \eqref{eq:generalQCQP} (one can then go as usual from an equality with $0$ constraint to two constraints of inequality with respect to $0$). We can also write $v_i^*v_i \geq 1$ rather than $v_i^*v_i = 1$, making usage of the fact that such a change does not alter whether the value of the program is $0$ or not. Also, while for ease of presentation we have stated the problem with $n$ variables, they can be easily combined into one single variable taking values in $\mathbb{C}^{n^2}$ in order to obtain a program of the exact same form as \eqref{eq:generalQCQP}. 

The number of such programs in dimension $n$ that we need to consider is given by the number of $n$-combinations with repetition out of a set of length $n$, equal to $2n-1 \choose n$. While asymptotically this will scale very quickly as a function of $n$, it will still be computationally tractable for values like $n=5$ or $n=6$, which goes beyond the theoretically understood range of up to $n=3$. To compute the final answer, one will take the minimum value out of all the programs. If this value is equal to zero, then the states $\{w_1, \ldots, w_n\}$  can be perfectly excluded with a projection, while if it is a non-zero value then perfect state exclusion of the set $\{w_1, \ldots, w_n\}$  will not be possible.

As for its applications to future results, there are two main consequences of the formalism we just described, beyond the indirect consequence of our work possibly inspiring further usage of the QCQP framework within quantum information processing.

The first of these consequences correspond to our newfound ability to use results about QCQPs in order to obtain new structural results about the perfect exclusion of pure states through projections. One can straightforwardly check that basic weak duality results will not help, since the value of the Lagrangian dual programs will always be zero. However, as we discussed earlier there is an ongoing stream of non-trivial theoretical results about QCQPs, and it seems reasonable to conjecture that some of those results will eventually apply to the highly structured programs that we consider.

The second of these consequences corresponds to the increased potential for the usage of standard mathematical optimization packages. While the work on solver software supporting QCQP is not yet at a stage giving a simple path for an implementation of the programs described by \eqref{eq:QCQP}, it seems reasonable to expect that such a stage will be reached in the near term. Then, such a piece of software could be compared with another one that implements the program in \eqref{eq:GeneralSDP}. From this, one would obtain a numerical study through standard solvers of the difference between POVMs and projections for perfect state exclusion of $n$ pure states in $n$ dimensions.

\subsection{Direct generalizations of our proof}

A naive approach for generalizing our result would start by considering conditions equivalent to the Caves-Fuchs-Schack inequality in the 4-dimensional case. However, this seems far from trivial, since the original derivation  in \cite{caves2002conditions} presents obstacles to such a generalization. In particular, it relies on the fact that when using the basis determined by an excluding projection, the sums corresponding to the inner products between two of the perfectly excluded states $\{a,b,c\}$ will have exactly one non-zero term. This makes it relatively easy to obtain formulas for the coefficients of $a$, $b$ and $c$ in that basis  as a function of the inner products between the states. However, solving the corresponding equations in $4$ dimensions seems like a significantly more complicated task, as each inner product between excluded states involves not $2$ but $4$ non-zero coefficients.

It could also be fruitful to take a geometrical perspective in order to better understand the situation at hand, following the approach in \cite{bengtsson2007geometry}. To see at an intuitive level what this might be like, one can start by observing that the space of density matrices is a section of the convex cone of positive semidefinite matrices. Also, the space of probability distributions with $3$ outcomes can be seen as an equilateral triangle, with each vertex of the triangle corresponding to a different deterministic distribution.   Then, as one can see in Chapter 10 of \cite{bengtsson2007geometry}, for any  fixed 3-outcome POVM the map which takes a density matrix to the probability distribution associated with applying the POVM to the density matrix  will be an affine map from the convex cone section to the equilateral triangle. 

In light of these facts, we can interpret any limits to state exclusion via projectors as saying that three points close to each other in the section of the convex cone cannot be sent to $3$ different faces of the triangle  by an affine map corresponding to a projection, as otherwise some points in the section would be sent outside the triangle, which is not allowed. Then, our result that projections are equivalent to POVMs can be seen as saying that in the case of pure states this does not change when we also allow the affine maps corresponding to non-projection POVMs. It might be interesting to fully formalize this thought, mathematically prove in this framework the known results about limits to state exclusion, and see if it is now easier to extend them to the case of $4$ pure states, where the space of outcomes of a $4$-outcome POVM can be seen as a regular tetrahedron. 

Another way in which a geometric perspective might useful would be for obtaining a constructive algorithm that transforms an excluding POVM into an excluding projection for the case we analyze in this paper ($3$ pure states in $3$ dimensions). It seems plausible that obtaining such an algorithm would then give insight about how to generalize our result.

Along the lines of using state exclusion characterizations alternative to the one given by \eqref{eq:GeneralSDP}, the work in \cite{1751-8121-51-36-365303} considers a generalization of the explicit perfect state exclusion criteria given in \cite{caves2002conditions} for the 2-dimensional case. However, it finds this generalization to be a sufficient condition for the $n$-dimensional case but not a necessary one. This work also observes that if pure states are perfectly excluded via a POVM, they will be an eigenvector (with eigenvalue $0$) of the corresponding POVM element, and they can be assumed to be an element of its spectral decomposition. Then, one can consider the feasibility of an optimization program where one tries to fill in  the remaining coefficients and vectors in the spectral decomposition of the POVM elements. While one might expect at first glance that the standard SDP framework in \eqref{eq:GeneralSDP} would offer a greater chance of applying mathematical optimization results, perhaps the fact that this formulation is closer in its shape to the QCQPs in  \eqref{eq:QCQP} could help make non-trivial connections between the projection case and the POVM case.

Note too that the main insight that leads to our result is the fact that one can take a POVM for perfect state exclusion to be an extremal one. This does not trivially lead to an answer, since there are extremal POVMs that are not projections, such as those in the family in Equations \eqref{firstMatrixEquation}--\eqref{thirdMatrixEquation}. However, an analysis of what the ranks of an extremal POVM in 3 dimensions have to look like allows us to obtain a parametrization of the situation that can be algebraically solved. The usage of more sophisticated facts about the structure of extremal POVMs (such as those facts derived in \cite{parthasarathy1999extremal, haapasalo2012quantum, sentis2013decomposition}) could be similarly involved in a generalization of our results to higher dimensionality. In fact, these considerations seem to us a very likely ingredient of any such generalization.

\subsection{Other considerations}

It might also be of interest to find relations between the optimality of projections for tasks involving POVMs, and the optimality of unitaries (without the use of ancillas) for certain tasks involving channels, discussed for example in \cite{beran2008optimality, arunachalam2013quantum}, specially considering the numerical evidence suggestive of such kind of connection identified in \cite{bandyopadhyay2014conclusive, arunachalam2013quantum}. When doing so, it might also be of interest to consider results (such as those in \cite{fefferman2016complete}) that characterize from a computational complexity point of view the power of computing with unitaries as opposed to general quantum channels.

Note too that for the case of mixed states, state exclusion is mathematically equivalent to state discrimination (the discriminated states would be those we obtain by computing $\I - \rho_i$). This means one can apply existing results about optimal measurements for mixed state discrimination \cite{eldar2004optimal, fiuravsek2003optimal}, and also consider the chance for generalizing results \cite{eldar2003designing, touzel2007optimal}  that look into the optimality of projections for state discrimination of pure states. Relatedly, as we discussed earlier the work in \cite{caves2002conditions} gives a characterization for perfect exclusion of 2-dimensional pure states that makes it clear that for any number of $k > 2$ states in 2 dimensions, projections are not necessarily equivalent to POVMs. It is the case that they additionally use a reduction to that setting to point out that for three mixed states in three dimensions, projections are not equivalent to POVMs within the context of perfect state exclusion.

Note as well that if one considers the possibility of constraining the number of non-zero components of a perfectly excluding POVM, the possibility of doing so simply corresponds to being able to perfectly exclude a subset of the set of states under consideration. Another related variation one might want to study is requiring that there are no zero components of a perfectly excluding POVM, as considered in \cite{1751-8121-51-36-365303}.

One could also look into determining whether the results here carry over to the gradual measure of PP-incompatibility defined in \cite{brun2015compatibility}, which is the value of the SDP in \eqref{eq:GeneralSDP} when the uniform distribution is assumed. This would correspond for example  to asking whether projections are optimal for state exclusion of 3 pure states in 3 dimensions even when perfect exclusion cannot be achieved by POVMs. If that was successfully answered, it would be natural to  relax assumptions even further, and consider arbitrary distributions. \cite{perry2016conclusive} offers a partial answer to these questions, by giving for an arbitrary number of pure states a sufficient condition for the existence of an optimal excluding POVM that is a projection (this is the condition that there is an optimal POVM such that none of the outcomes are perfectly excluded, and we also have that the pure states are linearly independent).

Note that the QCQP framework discussed in Section \ref{subsec:QCQP} extends without issues to the variants of the problem discussed in the previous paragraphs. In particular, if one wishes to study the exclusion of $k \neq n$ states, one can simply write $\{w_1, \ldots, w_k\}$  rather than $\{w_1, \ldots, w_n\}$ for the set of states to be excluded, giving rise to $k + n-1 \choose n$ $\( \text{rather than } {2n-1 \choose n} \)$ programs of the form in \eqref{eq:QCQP}. Similarly, if one wishes to study mixed states rather than pure states or introduce a probability distribution on the states, one can simply modify the objective function in \eqref{eq:QCQP} by  replacing the values of $s_i s_i^*$ with a corresponding $\rho_i$ and combining them with a multiplicative term $p_i$, respectively.

Furthermore, if one wishes to study non-perfect state exclusion, the programs given in Section  \ref{subsec:QCQP} can be used towards that purpose without additional modifications. As for the variant that limits the number of non-zero components of a perfectly excluding POVM, it will correspond to limiting the number of distinct terms that can appear in the $n$-combinations with repetition of $\{w_1, \ldots, w_n\}$ that characterize the programs in \eqref{eq:QCQP}. Similarly, the requirement that there are no zero components of the POVM corresponds to requiring that every state is excluded by a measurement outcome, and therefore to only considering the $n$-combination given by $\{s_1, \ldots, s_n\} = \{w_1, \ldots, w_n\}$.
\\

\begin{acknowledgments}

Thanks are due to John Watrous for numerous helpful suggestions, as well as to Juani Bermejo-Vega, Alex Bredariol-Grilo, Richard Cleve, Philippe Faist, Nicol\'as Guar\'in-Zapata, George Knee, Robin Kothari, Debbie Leung, Alexandre Nolin, Christopher Perry, Burak \c{S}ahino\u{g}lu, Jamie Sikora and Jon Tyson for insightful discussions. This work was partially supported by NSERC, the Canada Graduate Scholarship program, the Mike and Ophelia Lazaridis Fellowship program, and the David R. Chariton Graduate Scholarship program.
\end{acknowledgments}

\bibliographystyle{plainnat}
\bibliography{refsPP}

\end{document}